\documentclass[journal=mamobx,manuscript=article]{achemso}
\usepackage{graphicx}
\usepackage{amssymb}
\usepackage{amsmath}

\author{Yu Chai}
\affiliation{Molecular Foundry, Lawrence Berkeley National Lab, Berkeley, CA 94720, USA}
\author{Thomas Salez}
\affiliation{Univ. Bordeaux, CNRS, LOMA, UMR 5798, F-33405 Talence, France}
\alsoaffiliation{Global Station for Soft Matter, Global Institution for Collaborative Research and Education, Hokkaido University, Sapporo, Hokkaido 060-0808, Japan}
\author{James A. Forrest}
\affiliation{Department of Physics \& Astronomy,  University of Waterloo,  Waterloo, Ontario, N2L 3G1, Canada}
\alsoaffiliation{Perimeter Institute for Theoretical Physics, Waterloo, Ontario, N2L 2Y5, Canada}
\email{jforrest@uwaterloo.ca}

\title{Using $M_{\rm w}$ dependence of surface dynamics of glassy polymers to probe the length scale of free surface mobility}

\begin{document}
	
	\begin{abstract}
		We describe a series of surface levelling experiments in glassy polystyrene (PS) of varying molecular weight. The  evolution through a mobile surface layer is described by the glassy thin film equation that was introduced and used in a previous work\cite{Chai2014}. Excellent agreement with the data is achieved, with surface mobility as the single free parameter. Different molecular-weight dependencies in mobility are then observed above and below the glass transition. The results are discussed in terms of surface-chain anchoring in the bulk matrix, and the length scale for surface mobility.
	\end{abstract}
	
	\section{Introduction}
	
	The structure and dynamics of polymers in the near surface region close to the glass transition temperature ($T_{\rm g}$) have attracted much attention over the past decade because they have significant implications for polymer thin film fabrication, coating, and the lubrication industry\cite{Swallen2007}. In addition, from a fundamental science perspective, understanding how and why polymers behave differently at the surface, near $T_{\rm g}$, from those in the bulk is an important ongoing question\cite{Russell2017,Napolitano2017}. With more than 20 years of continuous research, accumulated evidence shows that the free surface of glassy polymers is not glassy and exhibits enhanced mobility compared to the same material in the bulk\cite{Fakhraai:2008Science, Yang:2010Science,Zuo2017,Zuo2013,Qi2013,Malshe2011a,Hoang2011,Kuon2018}.
	
	This enhanced mobility in the near surface region of glassy polymers is often attributed qualitatively and even quantitatively\cite{Chai2014, Yang:2010Science,Zhang2016} to the existence of a liquid-like layer near the surface. It has been widely (though not universally) accepted that this liquid-like layer is the boundary condition that leads to the decreases in the measured $T_{\rm g}$ observed in ultra-thin polymer films\cite{Forrest2001,Roth2005}. While theoretical and experimental\cite{Zuo2017, Forrest2013a, Forrest2014,Salez2015} efforts have been made to characterize this liquid-like layer, the nature of enhanced surface dynamics of glassy polymers is far from being fully understood. Direct measurements of depth dependent local relaxation rates with nanometer precision in depth have been preformed using $\beta$-NMR\cite{McKenzie2018}, but it is not clear how much these studies can be used to shed light on the $\alpha$-relaxation dynamics that is responsible for bulk flow.  Experimental studies of the surface relaxation of glassy films have been performed to provide a more quantitative picture\cite{Ilton2009,Teichroeb2003}, but none of those have yet been able to quantitatively provide the size of the liquid-like layer, $h^*$. However, we note that experiments\cite{Chai2014} and numerical simulations\cite{Tanis2019} on the surface levelling of glassy films provide a quantitative determination of the surface mobility $\frac{h^*}{3\eta}$, where $\eta$ is the viscosity of the liquid-like layer. A key step in improving our understanding of the surface properties of glassy polymers is to find a way to independently determine $h^*$ and $\eta$. 
	
	One possibility to isolate the size of the liquid-like layer, $h^*$, is to utilize the natural and tunable size of polymer molecules. This was done at a semi-quantitative level by Qi et al.\cite{Qi2013}, where nanoparticle embedding onto glassy PS surfaces as a function of the $M_{\rm w}$ of PS was studied. In that study, there was a distinct difference in the embedding behaviour between the large $M_{\rm w}$ values (87 kg/mol to 1200 kg/mol) and the low $M_{\rm w}$ values (3 kg/mol to 22 kg/mol). For the lowest $M_{\rm w}$ = 3 kg/mol (the one also used by Chai et al.\cite{Chai2014}), a surface flow was observed and is reminiscent of surface diffusion of organic glasses\cite{Daley2012b,Castez2009}.  This suggests that performing quantitative surface flow experiments on polymers with varying $M_{\rm w}$, and hence the size of the polymer molecules, may provide an independent estimate for $h^*$. Indeed, as the molecular size exceeds $h^*$, there may be quantitative changes in surface flow that allow for the determination of $h^*$. According to the previous study of Qi et al., there seems to be differences between the flow at the lowest value of $M_{\rm w}$ = 3 kg/mol, and that at larger $M_{\rm w}$ values, suggesting that the range from 3 kg/mol to 22 kg/mol should provide the transition in flow dynamics. This range also limits $M_{\rm w}$ to values less than the critical molecular weight (32 kg/mol for PS) as the entanglement of polymer chains can potentially affect surface properties\cite{Brown1996}.
	
	\section{Results and discussion}
	In this article, we describe a quantitative study of the levelling of PS stepped films (with varying $M_{\rm w}$) at temperatures above and below $T_{\rm g}$. By measuring the levelling profiles and fitting them to solutions of two different thin-film equations, we are able to distinguish without ambiguity whether the levelling is attributed to bulk or surface flow.

	\begin{figure}
		\includegraphics[width=0.7\linewidth]{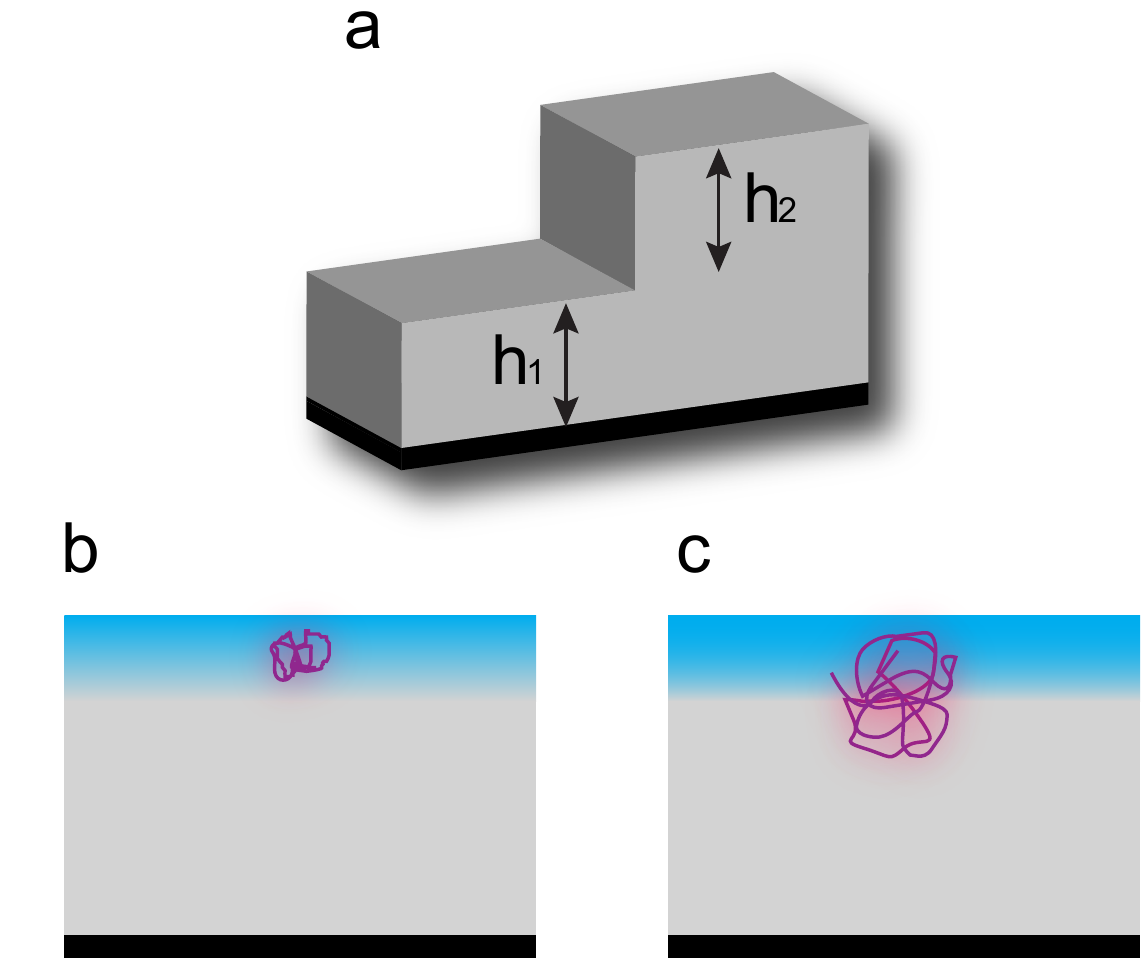}
		\caption{(a) Schematic diagram of a stepped polymer film, where the thickness of the top layer is denoted as $h_2$ and that of the bottom layer is denoted as $h_1$. (b,c) Some representative polymer chains near the surface: (b) 2$R_{\rm g}<h^*$: in this case, the entire polymer chain is in the liquid-like layer and surface flow is uninhibited; (c) 2$R_{\rm g}>h^*$:  while there are some polymer segments in the liquid-like layer, many parts of the chain are still in the glassy immobile region meaning surface flow is inhibited. }
		\label{fig:figure1}
	\end{figure}
	
	When the surface of a liquid film is not flat (and gravity can be neglected), surface tension drives a flow to minimize the surface area according to the equations of fluid dynamics.  Recently a number of studies have applied this idea to polymer stepped films, such as those shown in Fig. 1a. If the polymer is at a temperature $T>>T_{\rm g}$ then the flow is simply that of a thin liquid film. Such systems have been studied extensively and are well understood\cite{Stillwagon:1988JAP,Salez2012,McGraw:2011sm,McGraw:2012PRL}. In the case of glassy polymer films, bulk flow is not possible.  However, experimentally, it has been shown that the levelling of stepped films still occurs for glassy polymers.  In such cases, the levelling has been quantitatively described by a lubrication model involving the capillary-driven flow of a viscous layer localized at the free surface of the glassy film\cite{Chai2014}. 
	
	Mathematically, flows in both cases are described by two 2D $\rm{4^{th}}$ order partial differential equations. For the case where  $T > T_{\rm g}$, the equation is well studied, and we will refer to it as the Thin Film Equation (TFE)
	\begin{equation}
	\frac{\partial h}{\partial t} = \frac{\gamma}{3\eta}\frac{\partial}{\partial x}\left ( h^3\frac{\partial h}{\partial x}\right )
	\label{eqn:TFE}
	\end{equation}
	where $h(x,t)$ is the sample height profile at lateral position $x$ and time $t$, $\gamma$ is surface tension, and $\eta$ is bulk viscosity. This equation is non-linear, and is solved numerically for comparison with experiments\cite{Salez2012a}. For $T < T_{\rm g}$, flow occurs only through the liquid-like layer of thickness $h^*$. Because the latter is much smaller than the thickness of the film ($h^*<<h_1,h_2$), the equation is linear and can be solved analytically\cite{Chai2014,Salez2012}. This equation will be referred to as the Glassy Thin Film Equation (GTFE)
	\begin{equation}
	\frac{\partial h}{\partial t} = \frac{\gamma {h^*}^{3}}{3\eta}\frac{\partial^4 h}{\partial x^4}
	\label{eqn:GTFE}
	\end{equation}
	where now $\eta$ refers to the viscosity of the liquid-like layer. 
	
	The premise of using polymer chains to probe $h^*$ derives from the natural length scale of polymer molecules. In polymer melts, chains of a sufficient number of monomers can be described as random coils following configurations of an ideal random walk. While random chains do not have definitive configurations, we can define the statistical distribution of configurations for all polymer chains, which leads to ensemble averages of variables such as the end-to-end distance ($R_{\rm EE}$) and the radius of gyration ($R_{\rm g}$). The value of the ratio between $h^*$ and $R_{\rm g}$ (or $R_{\rm EE}$) can be expected to control the flow behaviour.  For example in Fig. 1b where $2R_{\rm g}<h^*$, the entire chain is in the liquid-like layer and thus surface flow is uninhibited.  In contrast, in Fig. 1c where $2R_{\rm g}>h^*$, while there are some polymer segments in the mobile region, the fact that many parts of the chain are in the glassy immobile region means that surface flow should be inhibited, and this may lead to differences in flow behaviour.
	
	Polystyrene (from Polymer Source Inc., PDI $<$ 1.1) with $M_{\rm w}$ = 3.0 kg/mol ($T_{\rm g}$ = 343 K), 11.9 kg/mol ($T_{\rm g}$ = 366 K), and 22.2 kg/mol ($T_{\rm g}$ = 369 K) were used in this study, where 2$R_{\rm g}$ are $\sim$ 3 nm, 6 nm, and 8 nm respectively. Although there is yet no consistent answer to how thick the liquid-like layer is, it has been reported by several groups that $h^*$ is only a few nanometers. Here we use the value reported by Paeng and co-workers\cite{Paeng2011a} as an estimation.  In their study, a temperature dependent $h^*$ of glassy PS was measured. Corresponding to our experimental temperatures, $h^*$ is 4 nm ($T$ = $T_{\rm g}$ - 6 K) to 6 nm ($T$ = $T_{\rm g}$ - 3 K). Given these estimations of $h^*$ and 2$R_{\rm g}$ for PS, and the study of Qi et al.\cite{Qi2013}, our range of $M_{\rm w}$ should enable us to probe the transition from uninhibited to inhibited flow. 
	
	Stepped PS films were made in a two-step process. First, polymer solutions in toluene were spin-coated onto two substrates, Si wafers with a size of 1 cm $\times$ 1 cm (University Wafer), and mica plates with a size of 2 cm $\times$ 2 cm. These spin-coated films were then annealed in a home-made oven flushed with dry nitrogen above $T_{\rm g}$ for more than 12 hours to remove internal stresses and residual solvents. The films were then slowly cooled back to room temperature. The thickness of the supported films on Si was measured by nulling ellipsometry, and that of mica supported films was measured by atomic force microscopy (AFM, JPK NanoWizard 3). Mica supported films were then transferred onto the surface of a clean water bath (Milli-Q). Typically, low $M_{\rm w}$ PS samples on water broke up  into small pieces that remained on the water surface. These film pieces were picked up by Si supported films. For the $M_{\rm w}=22$ kg/mol films, an initial cut with a razor blade was done on the mica supported films before floating as these films did not break by themselves during the transfer process. Finally, stepped films with many sharp steps ($h_{\rm 1}$ = $h_{\rm 2}$ = 90 nm, Fig. 1a) were dried for later study.
	
	All experiments were conducted on a JPK AFM with a heating stage, where the annealing temperature and duration were controlled by a Python script. The required precision on the film evolution meant it was necessary to keep track of the evolution of the same stepped front along time. Consequently, all measurements were conducted in series meaning that the levelling of only one stepped front was measured during one period of time. During the measurement, a sample was placed on the heating stage, and annealed at a predetermined temperature for a certain period of time. After each annealing cycle, the sample was cooled down to room temperature, and the height profile of the stepped front was measured by AFM in tapping mode. Time dependent height profiles were obtained at one temperature by repeating the annealing and imaging process until either (1) the experimental profile saturated, or (2) at least 90 hours of accumulated annealing time were reached. 
	
	\begin{figure}
		\includegraphics[width=1\linewidth]{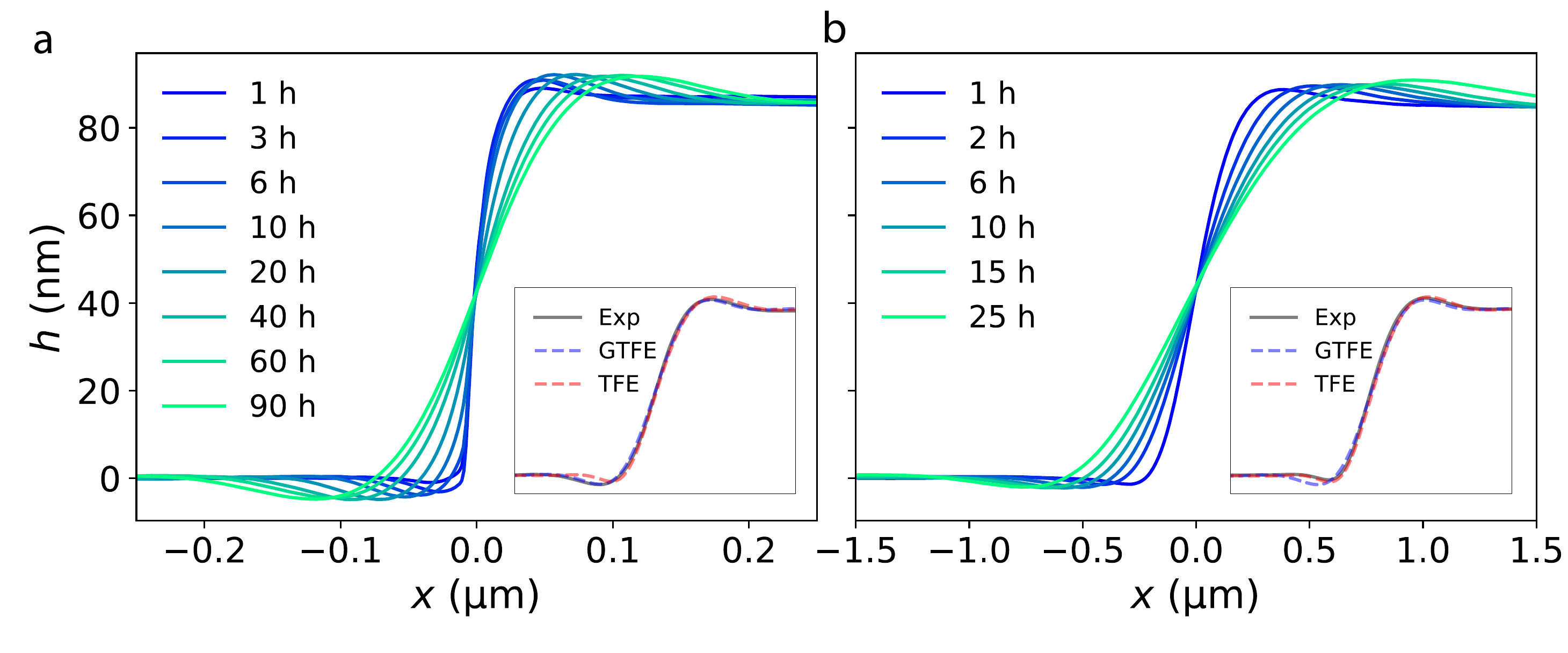}
		\caption{Temporal evolution of the stepped fronts for $M_{\rm w}$ = 11.9 kg/mol PS samples annealed at two temperatures: (a) $T$ = $T_{\rm g}$ - 6 K, (b) $T$ = $T_{\rm g}$ + 9 K. The insets show the experimental profiles (after 90 hours in (a), and 25 hours in (b)) compared with both the calculated TFE (Eq.~(\ref{eqn:TFE})) and GTFE (Eq.~(\ref{eqn:GTFE})) profiles.}
		\label{fig:figure2}
	\end{figure}
	
	Fig.2 shows typical evolutions of PS stepped  films of $M_{\rm w}$ = 11.9 kg/mol at temperatures below and above $T_{\rm g}$. In particular Fig. 2a shows the evolution at $T$ = $T_{\rm g}$ - 6 K, and Fig. 2b shows the evolution at $T$ = $T_{\rm g}$ + 9 K. There are two things revealed by these plots. First, $M_{\rm w}$ = 11.9 kg/mol PS at $T$ = $T_{\rm g}$ - 6 K exhibits some type of levelling below $T_{\rm g}$. Second, and less obviously, the shapes of the levelling profiles at two different annealing temperatures are not the same. This is analogous to our previous levelling study of $M_{\rm w}$ = 3.0 kg/mol PS, where we showed that above $T_{\rm g}$ the profile evolution proceeds by way of bulk film flow, while below $T_{\rm g}$ it proceeds by way of surface flow.
	
	In order to properly characterize the type of flow at play, we need to consider the detailed shapes of the levelling profiles. While it may not be clear from a casual inspection of the profiles in Fig. 2, we have previously developed a direct quantitative method to identify the flow mechanism. As long as the profiles have evolved sufficiently that the capillary surface stress can be approximated as $-\gamma \frac{\partial^2 h}{\partial x^2}$, and that the early-time residual and viscoelastic stresses have vanished, and if Eq.~(\ref{eqn:TFE}) (TFE) or Eq.~(\ref{eqn:GTFE}) (GTFE) is the governing equation, the profile $h(x,t)$ is in fact a function of the self-similar variable $xt^{-1/4}$ only. Furthermore, we can compare the experimental profile to both the TFE numerical solution\cite{Salez2012a}, and the GTFE analytical solution \cite{Salez2012}. This comparison is done quantitatively through our previously defined correlation functions: 
	\begin{equation}
	\chi_{\rm {GTFE}} = \frac{\int {\rm d}x(h_{\rm EXP}-h_{\rm TFE})^2}{\int {\rm d}x(h_{\rm GTFE}-h_{\rm TFE})^2}
	\label{chig}
	\end{equation}
	\begin{equation}
	\chi_{\rm {TFE}} = \frac{\int {\rm d}x(h_{\rm EXP}-h_{\rm GTFE})^2}{\int {\rm d}x(h_{\rm TFE}-h_{\rm GTFE})^2}
		\label{chit}
	\end{equation}
	where $h_{\rm EXP}$, $h_{\rm TFE}$, $h_{\rm GTFE}$ are the experimental, TFE, and GTFE profiles respectively. 
	Given an experimental profile, if $\chi_{\rm TFE}$ is equal to 1 and $\chi_{\rm GTFE}$ is equal to 0, the levelling process can be precisely described by the TFE (Eq.~\ref{eqn:TFE}). In other words, the entire film flows. In contrast, if $\chi_{\rm TFE}$ is equal to 0 and $\chi_{\rm GTFE}$ is equal to 1, the levelling process can be precisely described by the GTFE (Eq.~(\ref{eqn:GTFE})) implying a surface flow. 
	
	\begin{figure*}
		\includegraphics[width=1\linewidth]{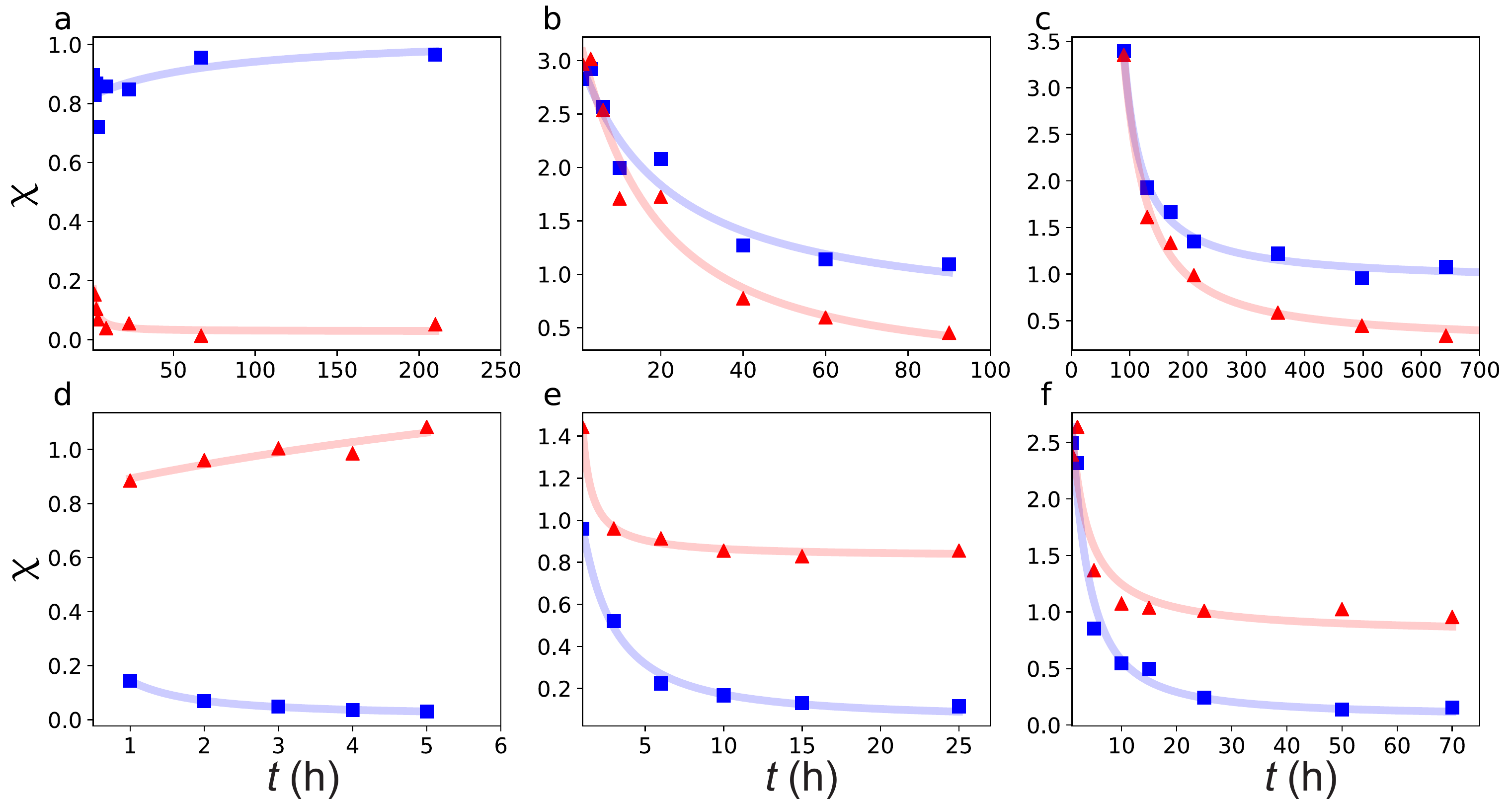}
		\caption{Temporal evolutions of the correlation functions (see Eqs.~(\ref{chig}) and~(\ref{chit})) $\chi$ for PS with $M_{\rm w}$ = 3.0 (a, d), 11.9 (b, e), 22.2 (c, f) kg/mol, where the blue squares stand for $\chi_{\rm GTFE}$, and the red triangles stand for $\chi_{\rm TFE}$. The annealing temperatures are (a) $T_{\rm g}$ - 5 K, (b, c) $T_{\rm g}$ - 6 K, (d) $T_{\rm g}$ + 10 K, (e, f) $T_{\rm g}$ + 9 K.}
		\label{fig:figure3}
	\end{figure*}

	Fig. 3 shows the previous time-dependent correlation functions for PS stepped films with three $M_{\rm w}$, annealed below or above $T_{\rm g}$. For $M_{\rm w}$ = 3 kg/mol PS (Fig. 3a), it is evident that both correlation functions, whether the sample is annealed above or below $T_{\rm g}$, reach their steady state values rapidly.  Furthermore, it is clear that for $T>T_{\rm g}$, $\chi_{\rm TFE}\simeq 1$  and $\chi_{\rm GTFE}\simeq 0$; while,  for $T<T_{\rm g}$, $\chi_{\rm TFE}\simeq 0$  and $\chi_{\rm GTFE}\simeq 1$. These results clearly indicate (as described previously \cite{Chai2014}) that for 3.0 kg/mol PS, the levelling process occurs through bulk flow for $T>T_{\rm g}$ and through surface flow for $T<T_{\rm g}$.
	
	For $M_{\rm w}$ = 11.9 (Fig. 3b and 3e) and 22.2 kg/mol PS (Fig. 3c and 3f), the behaviours of the correlation functions are novel and insightful. Perhaps the first thing to notice is that the values of all correlation functions are, at least initially, quite different from 0 or 1.  This is an indication that the profiles are not fit well by neither the GTFE nor the TFE models. While in Figs. 3e and 3f, where $T > T_{\rm g}$, the correlation functions do eventually converge to their expected values, in Figs. 3b and  3c where $T<T_{\rm g}$ this is not the case and the results remain transient. Note that in this case the profiles are still close to their sharp initial conditions, and the linear approximation of the curvature assumed in the derivations of Eqs.~(\ref{eqn:TFE}) and~(\ref{eqn:GTFE}) is certainly not valid. Specifically, for $M_{\rm w} = 11.9$ kg/mol and $T>T_{\rm g}$, after about 5-10 hours of annealing,  we have $\chi_{\rm TFE}\simeq 1$ and $\chi_{\rm GTFE}\simeq 0$ as expected for bulk flow.  For $T<T_{\rm g}$, after 90 hours of annealing, we have $\chi_{\rm GTFE}\simeq 1$, but $\chi_{\rm TFE}\simeq 0.5$.  While this is a fairly strong indication of surface flow, it does indicate that the 90 hour experimental time window was probably too short. For this reason, the total annealing time for the 22.2 kg/mol PS at $T<T_{\rm g}$ was increased to 650 hours (27 days). However, in this case again, $\chi_{\rm GTFE}\simeq 1$, but  $\chi_{\rm TFE}\simeq 0.5$.  While $\chi_{\rm TFE}$ does not reach its saturation value of 0 below $T_{\rm g}$ in the experimental time window for $M_{\rm w} = 11.9$ kg/mol and  $M_{\rm w} = 22.2$ kg/mol PS, it is safe to conclude from Fig. 3 that all PS samples show a bulk flow above $T_{\rm g}$, and surface flow or the onset of surface flow below $T_{\rm g}$. In addition, the levelling experiments for $M_{\rm w} = 11.9$ kg/mol PS stepped films at a temperature close to $T_{\rm g}$ show that the saturation of both correlation functions can be achieved in a reasonable annealing time at elevated temperatures (Fig. S1) or with thinner stepped films (Fig. S2), although how the interactions between thin polymer films and substrates affect the dynamics of the liquid-like layer still remains unclear. 
	
	Similar to the  conclusions by Qi et al.\cite{Qi2013}, that samples with $M_{\rm w}$ less than 22 kg/mol exhibit evidence for flow at temperatures below $T_{\rm g}$, all samples in the current study exhibit levelling, and thus flow, at $T= T_{\rm g}$ - 6 K or $T= T_{\rm g}$ - 5 K. As noted previously\cite{Chai2014}, as long as the system has evolved to the point where the evolution is self-similar, with $h(x,t)$ being a function only of the variable $xt^{-1/4}$, then the only adjustable parameter is the mobility $\frac{H^3}{3\eta}$ when fitting the solution of the TFE (Eq.~(\ref{eqn:TFE})) or the GTFE (Eq.~(\ref{eqn:GTFE})) to the experimental profile. Above $T_{\rm g}$, $H$ is equal to $h_1+h_2/2$ and $\eta$ is the bulk viscosity; while, below $T_{\rm g}$, $H$ is equal to $h^*$ and $\eta$ is the viscosity of the surface liquid-like layer. Note that, although for $M_{\rm w}=$ 11.9 and 22.2 kg/mol PS the correlation functions have not converged yet to their final values below $T_{\rm g}$, due to the extremely slow evolution and the fine profile details (bump, dip, etc.), the mobility values obtained from the fits are robust global features (essentially related to the evolution of the profile widths), and thus have saturated (see Fig. S1). Therefore, it is possible to extract PS mobility values for all our studied $M_{\rm w}$ and $T$, as shown in Fig.~4a. Since PS samples with different $M_{\rm w}$ values have different $T_{\rm g}$ values, the temperatures are normalized by using $T_{\rm g}/T$. As a comparison, the mobility values for $M_{\rm w} = $1.1, 1.7, and 3.7 kg/mol PS are calculated based on diffusion coefficients or viscosity values obtained in grating relaxation experiments by Zhang et al.\cite{Zhang2016}. Due to the use of different temperatures, the diffusion coefficients or viscosity values for $M_{\rm w} = $1.1, 1.7, and 3.7 kg/mol PS at two temperatures ($T_{\rm g}/T=$ 0.976 and 1.016) used here are extrapolated based on the fits presented in the same study by Zhang et al. Also shown as guides to the eye are power-law fits to the $M_{\rm w}$ dependence. Note that over this range we do not expect a single power law to be valid. Instead these fits are provided to demonstrate that the PS mobility below $T_{\rm g}$ shows a much stronger $M_{\rm w}$ dependence compared to that above $T_{\rm g}$. This implies that, for PS, surface mobility is more sensitive to the macromolecular size compared to bulk mobility. Fig. 4b further shows the ratio between the mobilities below $T_{\rm g}$ and above $T_{\rm g}$ as a function of $M_{\rm w}$. It is evident that the mobility ratio decreases significantly with $M_{\rm w}$, indicating a strong confinement effect due to the increase of the macromolecular size of the polymer chains. It is also noticeable that the mobility ratio shows a relatively weak $M_{\rm w}$ dependence for $M_{\rm w}>$ 11.9 kg/mol. This supports the idea of an intrinsic length scale less than the macromolecular size of $M_{\rm w}$ = 11.9 kg/mol PS. 
	
	\begin{figure}
		\includegraphics[width=0.9\linewidth]{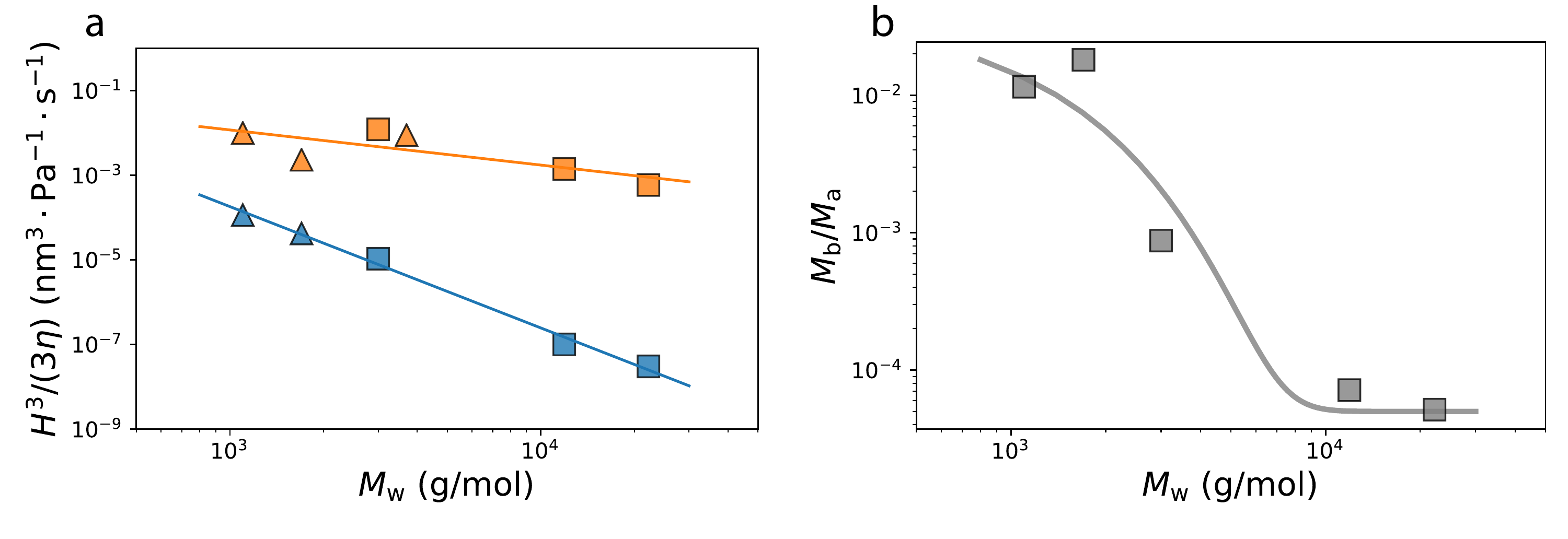}
		\caption{(a) Mobility of PS films as a function of molecular weight, for two rescaled temperatures (orange for $T_{\rm g}/T=0.976\pm 0.001$, and blue for $T_{\rm g}/T=1.016\pm 0.001$). Above $T_{\textrm{g}}$, $H=h_1+h_2/2$ and the bulk mobility is obtained by fitting the experimental profiles to the numerical solution of the TFE (Eq.(\ref{eqn:TFE})), while below $T_{\textrm{g}}$, $H=h^*$ and the surface mobility is obtained by fitting the experimental profiles to the analytical solution of the GTFE (Eq.(\ref{eqn:GTFE})). Our current results are represented by squares. Also shown are the results for 1.1 kg/mol PS, 1.7 kg/mol PS, and 3.7 kg/mol PS from Zhang et al.\cite{Zhang2016} (triangles). The solid lines represent power-law fits, as guides to the eye. (b) Ratio of the mobility values shown in (a) as a function of $M_{\rm w}$, where $M_{\rm b}$ represents the mobility at $T_{\rm g}/T=1.016\pm 0.001$, and $M_{\rm a}$ represents the mobility at $T_{\rm g}/T=0.976\pm 0.001$. The solid curve is a guide to the eye.}
		\label{fig:figure4}
	\end{figure}
	
	Before linking the different $M_{\rm w}$ dependences in PS mobility below $T_{\rm g}$ and above $T_{\rm g}$ to the ratio between $h^*$ and $2R_{\rm g}$, it is worthwhile examining the distribution of polymer conformations near a flat surface. For that purpose, we invoke a simple ideal random-walk based simulation.  In order to map the ideal random walks to real molecular conformations, we consider that for PS, one Kuhn monomer has a molar mass $M_{\rm 0}$ = 720 g/mol and one Kuhn length is $b$ = 1.8 nm\cite{Rubinstein2003}. We use the Kuhn length as the elementary spatial step size to ensure no correlation between two successive steps. $h^*$ is set to be three Kuhn lengths ($h^*=3b$ = 5.4 nm)\cite{Paeng2011a}. For each value of $M_{\rm w}$, and thus total chain length, we simulate ideal random walks that can possibly have a segment in the surface region. In order to accomplish this, each random-walk realization among a set of 10,000 attempts is started with one end at a distance $z=z_0$ from the free surface, and arrested after $N=M_{\textrm{w}}/M_0$ (in fact the integer part of it) steps. The value $z_0$ is then varied from 0 (the free surface) to $(N + 3)b$, since no chain starting at any $z_0$ greater than this last value can {\em possibly} have a segment in the liquid-like layer. While this ignores effects such as chain-end enrichment\cite{Jalbert1993a,Anastasiadis1988,Jang2000,Yethiraj1990}, and any enthalpic effects on chain conformations, it is a good starting point for our discussion.

	During the simulation, two variables are defined. These variables are $n_{\rm f}$, the number of free chains, and $n_{\rm g}$, the number of grafted chains in the liquid-like layer. We are interested in knowing what fraction of polymer molecules in the liquid-like layer can be considered as freely flowing. This can be alternatively rephrased as: ``What fraction of polymers that have {\it at least} one segment in the liquid-like layer have {\em all} segments in this layer?''. We illustrate this by an example in the inset of Fig. 5.  In the case (a), the entire polymer chain is in the liquid-like layer, i.e. this is a free polymer chain. For this realization, $n_{\rm f}$ is increased by 1. In the case (b), only  60\% of the polymer chain is in the liquid-like layer, i.e. this is a grafted polymer chain. Thus, $n_{\rm g}$ is increased by 0.6. In the case (c), no part of the polymer chain is in the liquid-like layer, i.e. this is an embedded polymer chain. Therefore, both $n_{\rm f}$ and $n_{\rm g}$ remain unchanged.
	
	\begin{figure}
		\includegraphics[width=0.9\linewidth]{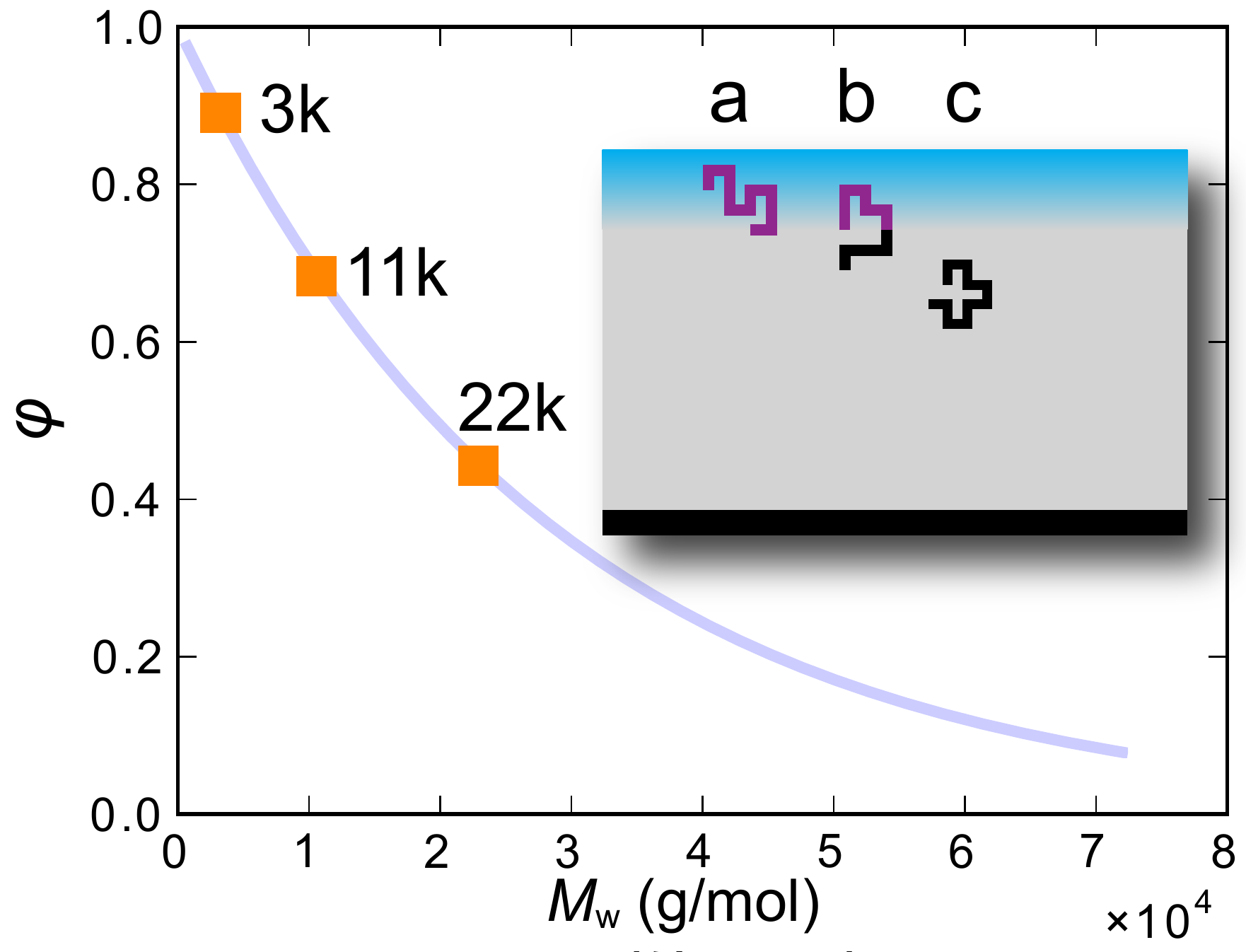}
		\caption{Ideal-random-walk simulated fraction of free polymers in the liquid-like layer as a function of PS $M_{\rm w}$. The inset shows three typical cases: (a) a free polymer chain, (b) a partially grafted polymer chain, (c) an embedded polymer chain.}
		\label{fig:figure5}
	\end{figure}

	After generating $10,000$ ideal random walks for each value of $z_0$ (among $N+3$ values), and for each value of $M_{\rm w}$, the fraction of free polymer chains in the liquid-like layer is calculated as
	\begin{equation}
	\varphi = \frac{n_{\rm f}}{n_{\rm f}+n_{\rm g}}.
	\end{equation}
	Fig. 5 shows the $M_{\rm w}$ dependent $\varphi$. Taking $M_{\rm w}$ = 3.0 kg/mol PS as an example, it gives $\varphi$ equal to 0.9 meaning that 90\% of the polymer chains in the liquid-like layer can flow. For $M_{\rm w}$ = 11.9 and 22.2 kg/mol PS, one get $\varphi$ values of 66\% and 46\% respectively. Based on these fractions, it is evident that there are many polymer chains mobile in the liquid-like layer, even when 2$R_{\rm g}$ $\geqslant$ $h^*$, but also a lot of grafted immobile chains. Although this structural picture of the liquid-like layer as a mixture of completely free and grafted chains can provide a simple explanation of the observed surface flow below $T_{\rm g}$ for the samples with 2$R_{\rm g}$ $\geqslant$ $h^*$, it is still not clear how to relate quantitatively the fractions obtained in Fig.~5 to the surface mobility in Fig.~4. Nevertheless, given the different molecular-weight dependencies in PS mobility above and below $T_{\rm g}$, it is possible to conclude that long-enough polymer chains in the liquid-like layer experience some confinement effect below $T_{\rm g}$. Thus, it is reasonable to deduce that the size of the liquid-like layer (assumed to be independent of $M_{\rm w}$) is comprised between 2 nm (2$R_{\rm g}$ of $M_{\rm w} = 1.1$ kg/mol PS) and 6 nm (2$R_{\rm g}$ of $M_{\rm w} = 11.9$ kg/mol PS).
	
	\subsection{Conclusion}
	We have measured by AFM the levelling process of PS stepped films with different $M_{\rm w}$, and found a tendency to surface flow below $T_{\rm g}$ in all samples. However, the different $M_{\rm w}$ dependencies in mobility observed above and below $T_{\rm g}$, and ideal-random-walk simulations, indicate that long-enough polymer chains in the liquid-like layer can experience a confinement effect. Thus, from our detailed analysis, we conclude that the size of the liquid-like layer should be comprised between 2 nm and 6 nm, and that surface flow for glassy PS films thicker than that might be inhibited for very large $M_{\rm w}$ values.
	
	The authors would like to acknowledge several helpful discussions with K. Dalnoki-Veress. Financial support from Natural Sciences and Research Council of Canada is gratefully acknowledged. Research at Perimeter Institute is supported by the Government of Canada through Industry Canada and by the Province of Ontario through the Ministry of Economic development and Innovation. JAF would like to thank ESPCI Paris for funding through the Paris Sciences Chair. 
	
	\bibliography{Molecular_weight_dependent_levelling}
\end{document}


\begin{figure}
		\includegraphics[width=0.7\linewidth]{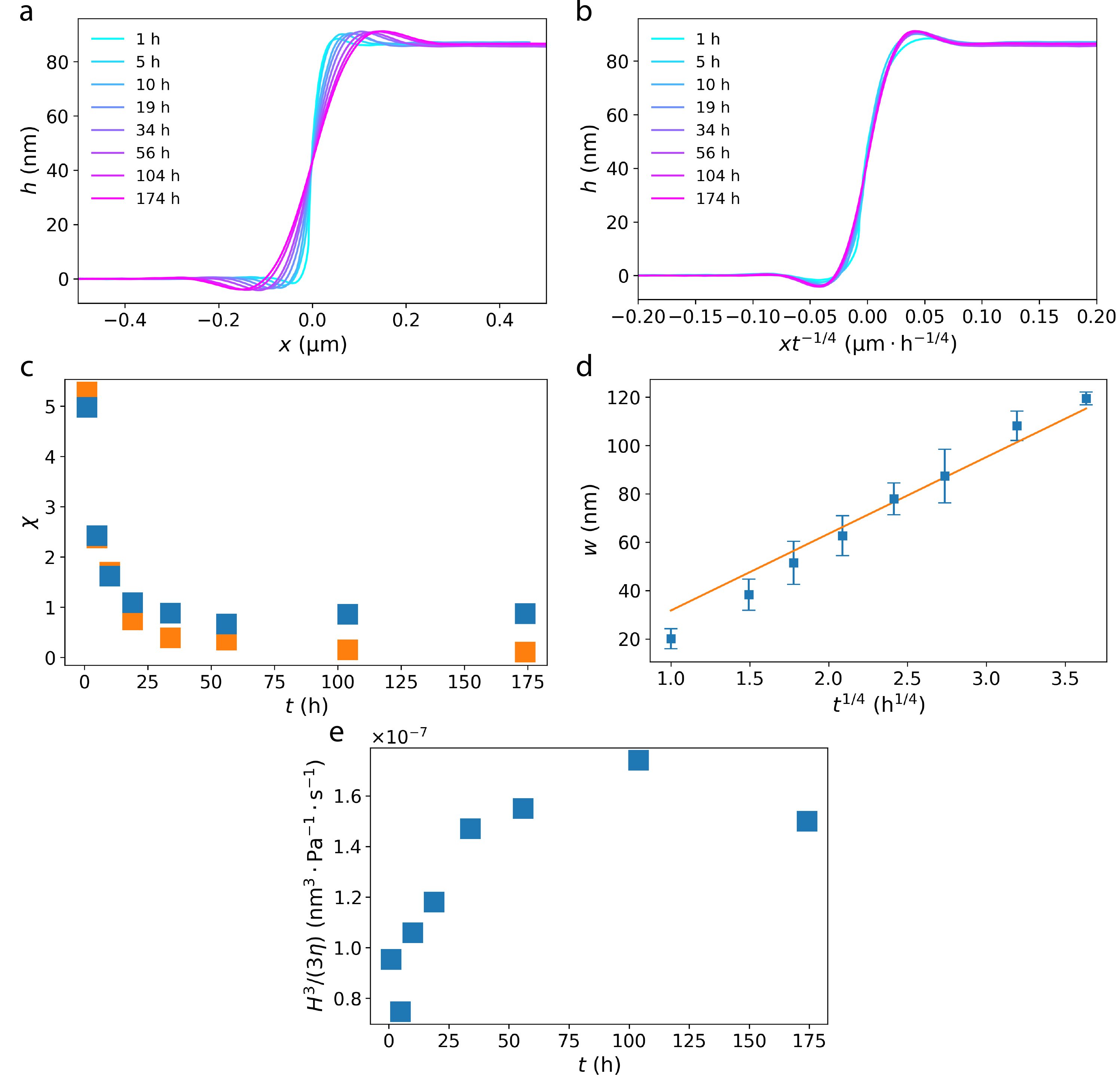}
		\caption{Levelling experiment of $M_{\rm w}$ = 11.9 kg/mol PS at $T=T_{\rm g}-$3 K, where $h_{\rm 1}=h_{\rm 2} = 90$ nm: (a) AFM height profiles at different times. (b) AFM height profiles at different times, with the self-similar variable $xt^{-1/4}$ as the new $x$-axis. (c) Time-dependent correlation functions introduced in the main text, where the blue squares stand for $\chi_{\rm GTFE}$ and the orange squares represent $\chi_{\rm TFE}$. (d) Time-dependent profile widths ($w$) determined by fitting the experimental profiles to ${\rm tanh}[x/(w/2)]$, where the solid line shows a $t^{1/4}$ relation. (e) Time-dependent surface mobility (where $H=h^*$) extracted from the fit of the self-similar experimental profile (panel b) to the solution of the GTFE (see main text).}
		\label{fig:figureS1}
	\end{figure}
    
    	\begin{figure}
		\includegraphics[width=0.7\linewidth]{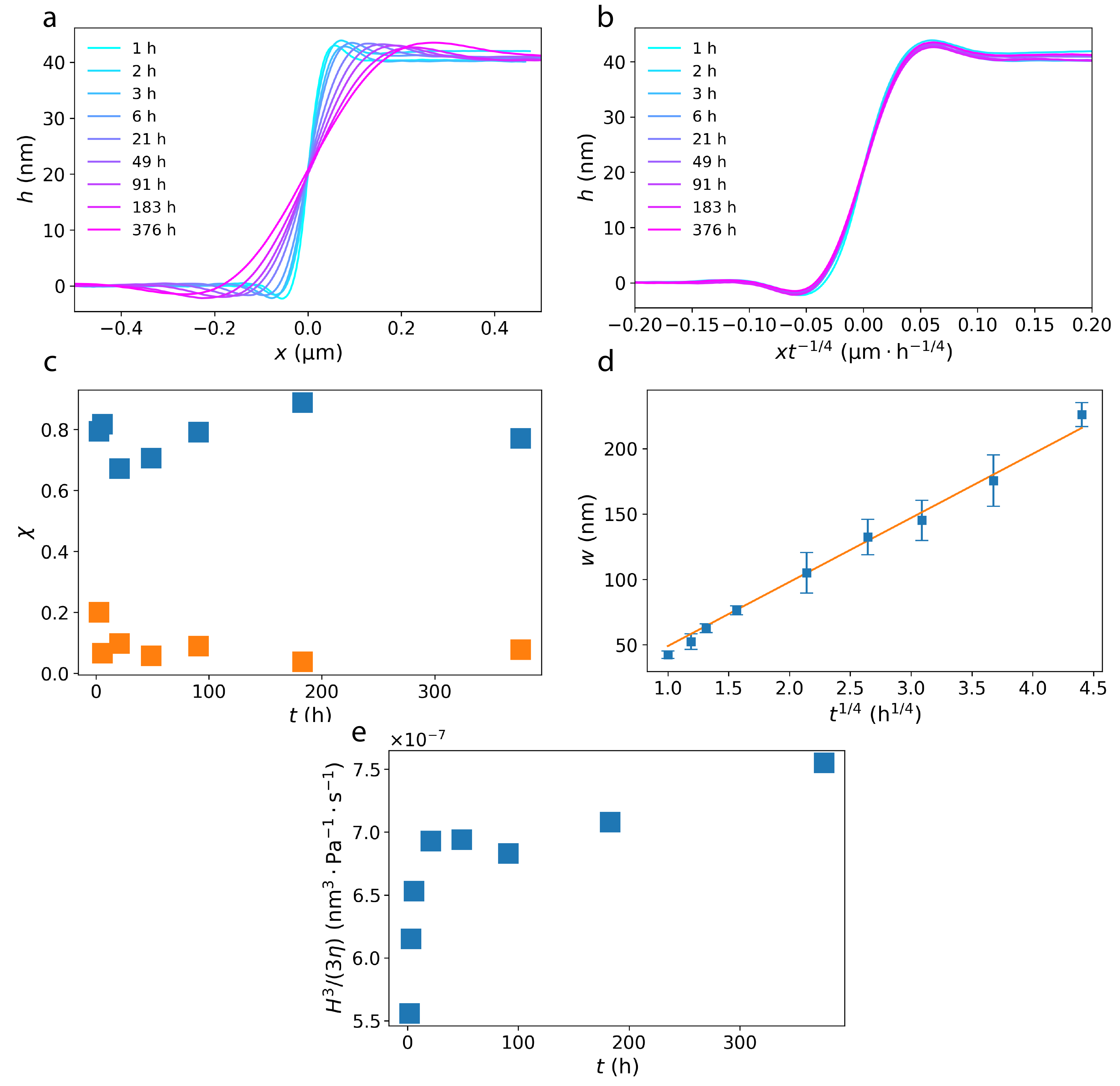}
		\caption{Levelling experiment of $M_{\rm w}$ = 11.9 kg/mol PS at $T=T_{\rm g}-$3 K, where $h_{\rm 1}=h_{\rm 2} = 40$ nm: (a) AFM height profiles at different times. (b) AFM height profiles at different times, with the self-similar variable $xt^{-1/4}$ as the new $x$-axis. (c) Time-dependent correlation functions introduced in the main text, where the blue squares stand for $\chi_{\rm GTFE}$ and the orange squares represent $\chi_{\rm TFE}$. (d) Time-dependent profile widths ($w$) determined by fitting the experimental profiles to ${\rm tanh}[x/(w/2)]$, where the solid line shows a $t^{1/4}$ relation. (e) Time-dependent surface mobility (where $H=h^*$) extracted from the fit of the self-similar experimental profile (panel b) to the solution of the GTFE (see main text).}
		\label{fig:figureS2}
	\end{figure}